\documentclass[twocolumn,groupedaddress,
superscriptaddress,
showpacs,preprintnumbers,
amsmath,amssymb,
aps,prl
]{revtex4-2}

\usepackage{graphicx}% Include figure files
\usepackage{dcolumn}% Align table columns on decimal point
\usepackage{bm}%
\usepackage[colorlinks=true,citecolor=blue]{hyperref}
\hypersetup{colorlinks=true,citecolor=blue,linkcolor=red,urlcolor=blue}

\usepackage[T1]{fontenc}
\usepackage{float}

\usepackage{color}
\definecolor{Green}{RGB}{0,204,102}
\definecolor{Purple}{RGB}{102,0,255}
\definecolor{Blue}{RGB}{51,153,255}
\definecolor{Red}{RGB}{151,010,010}

\begin{document}

\sloppy

\title{Electron correlations rule phonon-driven instability in single layer TiSe$_2$}

\newcommand*{\DIPC}[0]{{
Donostia International Physics Center (DIPC), 20018 Donostia-San Sebasti\'an, Spain}}

\newcommand*{\IFS}[0]{{
Institute of Physics, 10000 Zagreb, Croatia}}

\newcommand*{\zahra}[0]{{
School of Nano Science, Institute for Research in Fundamental Sciences (IPM), 19395-5531 Tehran, Iran}}

\newcommand*{\ivor}[0]{{
Ru\dj er Bo\v{s}kovi\'c Institute, 10000 Zagreb, Croatia}}

\author{Dino Novko}
\email{dino.novko@gmail.com}
\affiliation{\IFS}
\affiliation{\DIPC}
\author{Zahra Torbatian}
\affiliation{\zahra}
\author{Ivor Lon\v{c}ari\'c}
\affiliation{\ivor}

\begin{abstract}
We investigate the controversial case of charge-density-wave (CDW) order in single layer 1T-TiSe$_2$ by employing the density functional perturbation theory with on-site Hubbard interactions. The results emphasize the crucial role of electron correlations via Hubbard corrections in order to capture the accurate electronic structure, low- and high-temperature limits of the CDW phonon mode, and temperature-charge phase diagram. We show, in close agreement with the experiments, that total phase diagram consists of both commensurate and incommensurate CDW regions, where the latter coincide with the superconductive phase and might be instrumental for its formation.
In addition to the established roles of quantum lattice fluctuations and excitonic interactions, our analysis emphasizes the overlooked crucial role of the momentum dependent electron-phonon coupling and electron correlations for the CDW phase transition in single layer TiSe$_2$.

%According to our analysis, the CDW order in TiSe$_2$ is mainly governed by momentum dependent electron-phonon coupling and electron correlations, while other mechanisms, such as exchange and excitonic interactions, might only support its formation.

%Transition metal dichalcogenides host a variety of unconventional ordered and superconductive phases central to the condensed matter physics.

%We investigate the controversial case of charge-density-wave (CDW) order in single layer 1T-TiSe$_2$ by employing the density functional perturbation theory with on-site Hubbard interactions. The results emphasize the crucial role of electron correlations via Hubbard corrections in order to capture the accurate electronic structure, low- and high-temperature limits of the CDW phonon mode, and temperature-charge phase diagram, all overestimated with semi-local functionals. According to our analysis, the CDW order in TiSe$_2$ is mainly governed by momentum dependent electron-phonon coupling and electron correlations, while other mechanisms, such as exchange and excitonic interactions, might only support its formation.

\end{abstract}

\maketitle

%\section*{Introduction}\label{sec:intro}

Despite almost five decades of extensive research\,\cite{gruner88,rossnagel11}, microscopic origins of charge-density-wave (CDW) formation in bulk and two-dimensional (2D) 1T-TiSe$_2$ are still a matter of active debate\,\cite{otto21,duan21,cheng22,zhang22}, where either electron-phonon\,\cite{yoshida80,rossnagel02,calandra11,karam18,zhou20,wegner20,watson20} or electron-electron\,\cite{disalvo76,traum78,cercellier07,monney09,rohwer11,kogar17,chen18,lian19} interactions are discussed as a main driving force.

%Layered transition metal dichalcogenides host a variety of unconventional ordered and superconductive phases central to the condensed matter physics and chemistry\,\cite{gruner88,rossnagel11}. Despite almost five decades of extensive research, microscopic origins of these phase transitions are still a matter of active debate. Particularly controversial is a mechanism behind the charge-density-wave (CDW) formation in bulk and two-dimensional (2D) 1T-TiSe$_2$, where either electron-phonon\,\cite{yoshida80,rossnagel02,calandra11,karam18,zhou20,wegner20} or electron-electron\,\cite{disalvo76,traum78,cercellier07,monney09,rohwer11,kogar17,chen18,lian19} interactions are discussed as a main driving force.

In phonon-driven scenario\,\cite{hughes77,whangbo92} relevant phonon mode is softened by electron-hole pair transitions between center and edge of the Brillouin zone (BZ)\,\cite{yoshida80}, which pushes the system in new ground state with distorted bonds. A key role of electron-phonon coupling (EPC) for CDW in TiSe$_2$ was corroborated by several experimental observations, namely, temperature dependent Kohn anomaly\,\cite{holt01,weber11}, large periodic lattice distortions (PLD)\,\cite{fang17}, high electrical resistivity\,\cite{disalvo76,knowles20}, and superconductivity at higher pressures and by Cu intercalation\,\cite{morosan06,qian07,zhao07,kusmartseva09}. 
%However, a proper microscopic support given by means of quantitative theoretical approaches is still missing.
Even though density functional theory (DFT) is able to reproduce a correct PLD and phonon frequencies of the low-temperature CDW structure, as well as strong EPC\,\cite{calandra11,hellgren17,zhou20}, it fails in explaining the correct transition temperature $T_{\rm CDW}$, the temperature dependence of the CDW mode both for bulk and monolayer (1L) as well as the charge melting of their CDW phases\,\cite{duong15,fu16,wei17,singh17,guster18,chen18}, and, therefore, does not provide a proper microscopic support for the phonon-induced instability.

In purely electron-driven, or excitonic insulator instability\,\cite{keldysh65,jerome67} stabilization of the charge order comes from the soft electronic mode, i.e., exciton or plasmon, which was confirmed to exist in TiSe$_2$ by means of the electron loss spectroscopy\,\cite{kogar17} and later by the DFT calculations\,\cite{lian19}. Prerequisites for this spontaneous exciton condensation are believed to be small indirect or negative energy gap and weakly screened electron-hole interaction. Recent angle-resolved photoemission (ARPES) study claimed that purely excitonic mechanism is unlikely in 1L TiSe$_2$\,\cite{watson20}, while the presence of ungapped Fermi surface for CDW phase in bulk might quench the exciton and downsize its role\,\cite{watson19}.

There are, in fact, growing evidences, extracted from theoretical models\,\cite{wezel10,watanabe15,kaneko18,lian20} and various time-resolved pump-probe experiments\,\cite{mathias16,monney16,porer14,karam18,hedayat19,burian21,duan21,otto21,cheng22}, that both subsystems cooperate to induce the CDW in TiSe$_2$. It is, therefore, paramount to construct coherent and quantitative microscopic theory that can single out a dominant driving force.

%There are, in fact, growing evidences that both subsystems cooperate to induce the CDW in TiSe$_2$\,\cite{kidd02,hedayat19}. Various theoretical models\,\cite{wezel10,vanwezel10,watanabe15,kaneko18,lian20} and powerful time-resolved pump-probe experiments\,\cite{rohwer11,mohrvorobeva11,hellmann12,mathias16,monney16,porer14,karam18,hedayat19,hedayat21,burian21,duan21,otto21,cheng22} were utilized to resolve whether phonons or electrons dominate the process. However, it is unclear to what extent ultrafast techniques can be used to comprehend the ground state CDW, since the corresponding time scales are often inseparable\,\cite{baranov14}, while many-body interactions are modified under non-equilibrium regime\,\cite{hu22}.

%All these leaves the question of the origin of the CDW in TiSe$_2$ still open.

%Alternatively, quantitative and careful theoretical methods should aid us in order to reach this goal.

In order to reach the aforesaid goal, we investigate 1L TiSe$_2$ and provide reliable evidences showing that 2D charge order is governed by the unconventional phonon-driven instability, underlain by the momentum-dependent EPC, as, e.g., it is the case in NbSe$_2$\,\cite{johannes06,calandra09,zhu15}, as well as significant electron correlations, which are crucial for obtaining the right transition temperatures $T_{\rm CDW}$ and the temperature-charge phase diagram, consisting of commensurate and incommensurate CDW regions.
%The melting of the CDW state, although experimentally relevant, is still not properly explained.
Accordingly, invoking the presence of purely electronic soft modes in the context of CDW melting or formation is not crucial in single layer TiSe$_2$. The latter is in line with the recent time-resolved ultrafast electron diffraction study of bulk TiSe$_2$ indicating that excitonic correlations are in charge only for the out-of-plane CDW order\,\cite{cheng22}.

To simulate phonon dynamics while properly accounting for electron correlations, we utilize density functional perturbation theory (DFPT)\,\cite{baroni01,qe} corrected with on-site Coulomb (Hubbard) interaction U\,\cite{floris11,timrov18,floris20} (see Supplemental Material (SM)\,\cite{SM} for more computational details), which was proven to be affordable and accurate tool for quantitative studies of EPC in correlated materials\,\cite{zhou21}.

%%%%%%%%%%%%%%%%%%%%%%%%%%%%%
%%%%%%%%%%%%%%FIGURE%%%%%%%%%
%%%%%%%%%%%%%%%%%%%%%%%%%%%%%
\begin{figure}[!t]
\begin{center}
\includegraphics[width=\columnwidth]{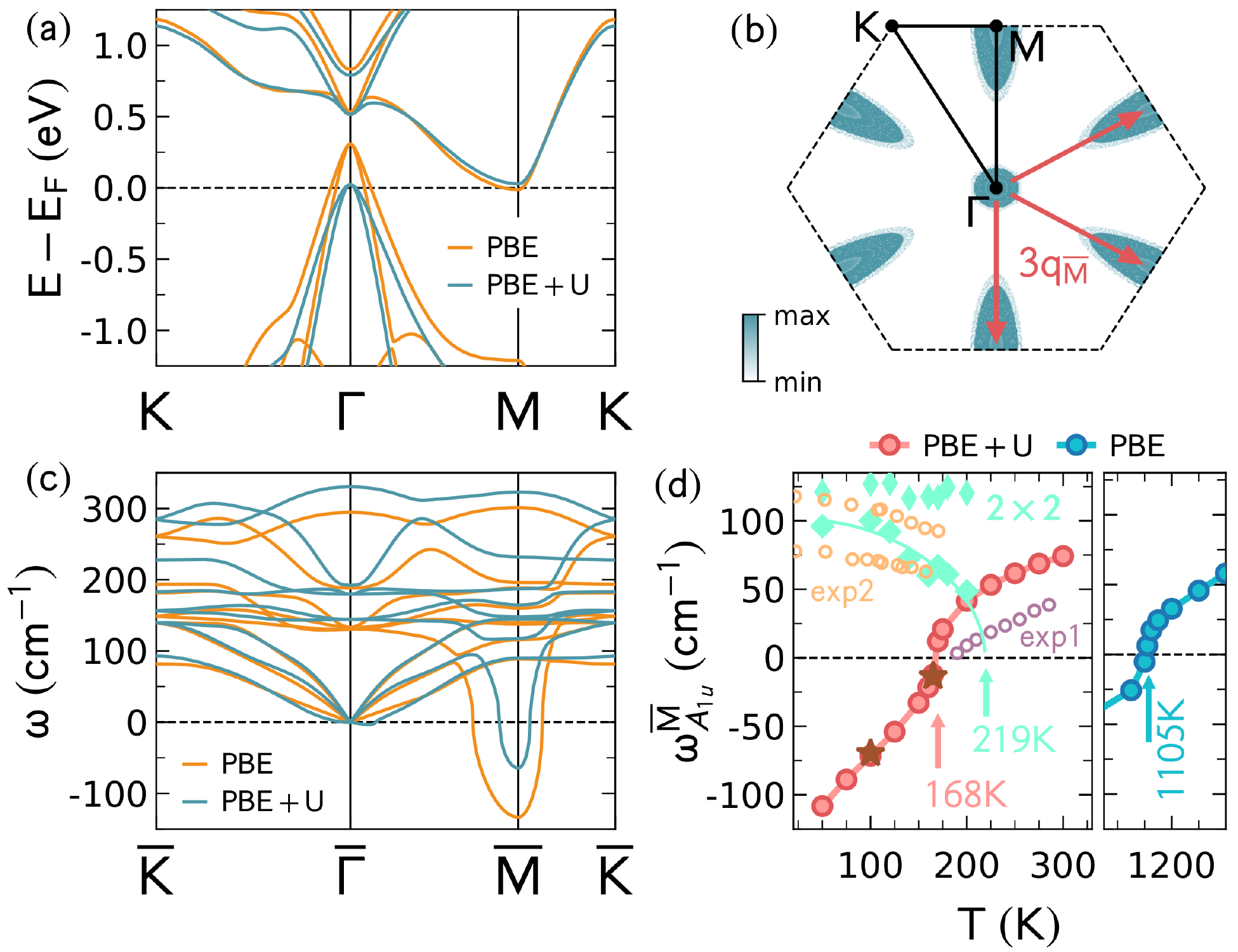}
\caption{(a) Electronic band structures of the TiSe$_2$ monolayer along $\mathrm{K}-\Gamma-\mathrm{M}-\mathrm{K}$ symmetry path obtained by DFT calculations with (PBE$+$U, blue) and without (PBE, orange) Hubbard correction. Hubbard parameter is ${\rm U}=2.98$\,eV and electronic temperature $T=350$\,K. (b) Fermi surface obtained with PBE$+$U. Red arrows mark the three nonequivalent electron-hole pair transitions. (c) Phonon band structures along $\mathrm{\overline{K}}-\overline{\Gamma}-\mathrm{\overline{M}}-\mathrm{\overline{K}}$ path obtained with PBE and PBE+U at $T=100$\,K. Negative frequencies represent imaginary values. (d) Corresponding frequencies of the soft CDW mode at the $\mathrm{\overline{M}}$ point as a function of electron temperature. For the low-$T$ $2\times2$ phase the results of the two CDW amplitude modes are shown with aqua diamonds. Red and blue circles are obtained with $\mathbf{k}=96\times96\times1$ momentum grid, while brown stars with $\mathbf{k}=120\times120\times1$. Experimental results for the high-$T$ and low-$T$ phases are from Refs.\,\cite{holt01} and \cite{sugai80}, respectively.
}
\label{fig1}
\end{center}
\end{figure}
%%%%%%%%%%%%%%%%%%%%%%%%%%%%%

Figure \ref{fig1}(a) shows the electronic band structure of 1L TiSe$_2$ along high symmetry points obtained by using semi-local exchange-correlation PBE functional corrected with Hubbard on-site interaction (PBE$+$U), compared with pure PBE result. Here we use Hubbard on-site energy of ${\rm U}=2.98$\,eV for Ti atom, obtained self-consistently by using a first-principles linear response approach\,\cite{timrov18}. PBE band structure is metallic with strongly depopulated Se-$4p$ hole-like states at the $\Gamma$ point and partially occupied Ti-$3d$ electron-like states at the M point\,\cite{calandra11,hellgren17,singh17,guster18}, at odds with ARPES measurements\,\cite{chen15,sugawara16,jia18,watson20}. By including Hubbard correction the Se-$4p$ bands at $\Gamma$ become almost fully populated (see also Fig.\,S1 in SM\,\cite{SM}), while keeping electron-like bands at M point intact. Therefore, electron correlations transform 1L TiSe$_2$ to an almost zero-gap semiconductor, in agreement with early measurements\,\cite{sugawara16}, and hybrid-functional calculations, which include portion of the exact exchange\,\cite{chen15,hellgren17,zhou20}. Note that further quasiparticle renormalizations due to electron-phonon\,\cite{ortenzi09} and electron-hole\,\cite{monney15} interactions might slightly open the gap between electron- and hole-like bands, which would be in accordance with recent ARPES study\,\cite{watson20}. Further analysis (see Fig.\,S1\,\cite{SM}) demonstrates considerable sensitivity of low-energy bands on excess charge concentration, temperature, and value of U, which is relevant for apprehending the mechanism of the CDW formation. For instance, low (high) temperature regime is governed by electron-like (hole-like) carriers, which is known from the transport experiments\,\cite{watson19b,knowles20}.
All these re-confirms the already established role of U corrections for obtaining the right electronic structure of both 1L and bulk TiSe$_2$\,\cite{bianco15,chen16,lian19,lian20}.

Fermi surface of the PBE$+$U band structure [Fig.\,\ref{fig1}(b)] already indicates that perfect nesting, a precondition for purely electronic, i.e., Peierls instability, is not probable in TiSe$_2$\,\cite{disalvo76,rossnagel02}. Nevertheless, three nonequivalent interband transitions from the center to the edge of the BZ are possible and instrumental for the appearance of the Kohn anomaly and formation of the PLD.

Phonon band structures of 1L TiSe$_2$ with and without proper description of electron correlations are depicted in Fig.\,\ref{fig1}(c). Inclusion of Hubbard correction blueshifts most of the phonon spectra, which could be explained as the overall reduced screening, i.e., transition from metal to small-gap semiconductor when the correct U is added. Particularly relevant is the hardening of the $A_{\rm 1u}$ acoustic phonon frequency at $\mathbf{q}=\mathrm{\overline{M}}$, unstable mode which is believed to be instrumental for the CDW formation, accompanied with characteristic PLD and opening of the charge-order gap, in the phonon-induced scenario\,\cite{rossnagel11,zhou20,watson20}. Microscopic origins of this electron-correlation-induced phonon renormalization are analyzed below. Consequently, the CDW transition (electron) temperature is reduced from $T_{\rm CDW}=1105$\,K, as obtained with DFPT calculations and PBE functionals, to $T_{\rm CDW}=168$\,K [see Fig.\,\ref{fig1}(d)], greatly improving the agreement with the experiments\,\cite{peng15,chen15,chen16,sugawara16,kolekar18,jia18}. Softening and hardening of the CDW phonon as a function of electron temperature for normal (high-$T$) and CDW (low-$T$) phases also reproduces nicely the trends as observed in the inelastic x-ray\,\cite{holt01,weber11} and Raman\,\cite{sugai80} studies.
Note that the amplitude mode softening for the low-$T$ $2\times 2$ reaches zero for $T_{\rm CDW}\approx 220$\,K. This comes from additional relaxation of the Se atoms with PBE+U and for each of the temperatures, while the previously mentioned result ($T_{\rm CDW}=168$\,K) is obtained for fixed Se atoms and relaxed with PBE for $T\ll T_{\rm CDW}$ (e.g., distances between the Ti and Se planes are $1.545$\,\AA~ and $1.563$\,\AA~ for PBE and PBE+U). This emphasizes the sensitivity of the CDW properties with respect to length of the Ti-Se bonds, as previously discussed\,\cite{zunger78,fu16}.

According to the previous DFT-based studies\,\cite{hellgren17,zhou20} both PBE and hybrid functionals significantly overestimate the value of $T_{\rm CDW}$ in 1L and bulk TiSe$_2$, while the inclusion of the long-range exchange via HSE hybrid functional enhances the EPC of the bulk CDW soft mode. Having in mind our findings, this leads to the following hierarchy of the calculated phonon frequencies $\omega^{\rm PBE+U}_{\rm CDW}<\omega^{\rm PBE}_{\rm CDW}<\omega^{\rm HSE}_{\rm CDW}$, and the corresponding transition temperatures $T^{\rm PBE+U}_{\rm CDW}<T^{\rm PBE}_{\rm CDW}<T^{\rm HSE}_{\rm CDW}$\,\cite{hellgren17,zhou20}. If one assumes that nesting is not significantly modified in these three approaches (as it was shown at least to be the case for PBE and HSE\,\cite{zhou20}), one might draw the conclusion that EPC strengths of the CDW soft mode at $\mathbf{q}=\mathrm{\overline{M}}$ are ordered as $\lambda^{\rm PBE+U}<\lambda^{\rm PBE}<\lambda^{\rm HSE}$ (see Figs.\,S2 and S3 in SM for additional comparison between the three functionals\,\cite{SM}).

We would also like to point out that structural stability and lattice dynamics of bulk TiSe$_2$ as obtained with DFT$+$U  was carried out in Refs.\,\cite{bianco15} and \cite{hellgren17}, where it was concluded that the CDW instability is suppressed for certain values of U that reproduce the correct electronic structure. However, it was further discussed that these contradictory results might be underconverged and that certain improvements are necessary\,\cite{bianco15} (see also Sec.\,S2 in SI). In addition, quantum lattice fluctuations were shown to be important to understand the CDW transitions in various transition metal dichalcogenides\,\cite{zhou20,leroux15,zheng22}. For instance, in the case of TiSe$_2$ monolayer, it reduces the harmonic PBE transition temperature from $1195$\,K to $440$\,K\,\cite{zhou20}. Here we suggest that anharmonic corrections (which are always present in the case of CDW transitions) depend on the chosen functional, and, therefore, have different impact in the case of PBE and PBE$+$U (see Fig.\,S4 in SM\,\cite{SM}). Also, as we show below, the right electronic structure as obtained with PBE$+$U is essential for reproducing certain CDW features (e.g., melting of the CDW with electron doping and incommensurate CDW) that cannot be explained in terms of phonon entropy.

%Here we suggest that anharmonic corrections should be applied on top of PBE$+$U in order to test its role, since already present PBE$+$U calculations provide electronic structure and phonon frequencies in a very close agreement with the experiments.

Three $\mathbf{q}=\mathrm{\overline{M}}$ soft phonon modes with nonidentical (but symmetry related) displacement patterns (induced by previously mentioned three $\Gamma\rightarrow \mathrm{M}$ interband transitions) are responsible for the well-established PLD and $2\times 2$ CDW supercell\,\cite{disalvo76,guster18}, shown in Fig.\,S5 of SM\,\cite{SM}. For the low-$T$ structure, the Ti-$d$ bands are folded from the edge to the center of the BZ, where two out of three unoccupied conduction states hybridize with the Se-$p$ states forming a CDW gap, while the lowest one remains unperturbed.
Experimentally determined gap between the second conduction Ti-$d$ and occupied Se-$p$ bands at $\Gamma$ in the CDW phase was reported to be in the $0.35-0.4$\,eV range\,\cite{li07,monney12}. Here we obtain the values of $\sim 0.15$\,eV with PBE$+$U and $\sim 0.62$\,eV with PBE. Infrared spectroscopy study showed that the low-energy edge of the CDW gap extends towards 0.15\,eV\,\cite{li07}, while the smallest gap between occupied and second unoccupied bands slightly away from $\Gamma$ are $\sim 0.1$\,eV and $\sim 0.4$\,eV for PBE$+$U and PBE, respectively. This shows how PBE$+$U provides a good electronic structure even for the low-$T$ $2\times2$ distorted structure.
Regarding the calculated PLD in the low-$T$ regime, the displacements of the Ti and Se atoms obtained with PBE are in very good agreement with x-ray diffraction experiments\,\cite{fang17}, while PBE$+$U provides almost five times smaller values for Ti atoms (see Fig.\,S6\,\cite{SM}). Within the present approach, this might indicate that besides the dominant EPC mechanism, there are other interactions supporting the CDW formation, like exciton-phonon coupling that was shown to increase PLD\,\cite{kaneko18}.

Note also the following consistent pattern: in comparison with PBE calculations Kohn anomaly obtained with PBE$+$U is smaller, transition temperature $T_{\rm CDW}$ is lower, PLD and the concomitant CDW gap are reduced (see also Fig.\,S3\,\cite{SM}).
The goal of the following analysis is to decipher in more details microscopic origins underlying the Kohn anomaly of the CDW mode and the above-mentioned pattern.

%%%%%%%%%%%%%%%%%%%%%%%%%%%%%
%%%%%%%%%%%%%%FIGURE%%%%%%%%%
%%%%%%%%%%%%%%%%%%%%%%%%%%%%%
\begin{figure}[!t]
\begin{center}
\includegraphics[width=\columnwidth]{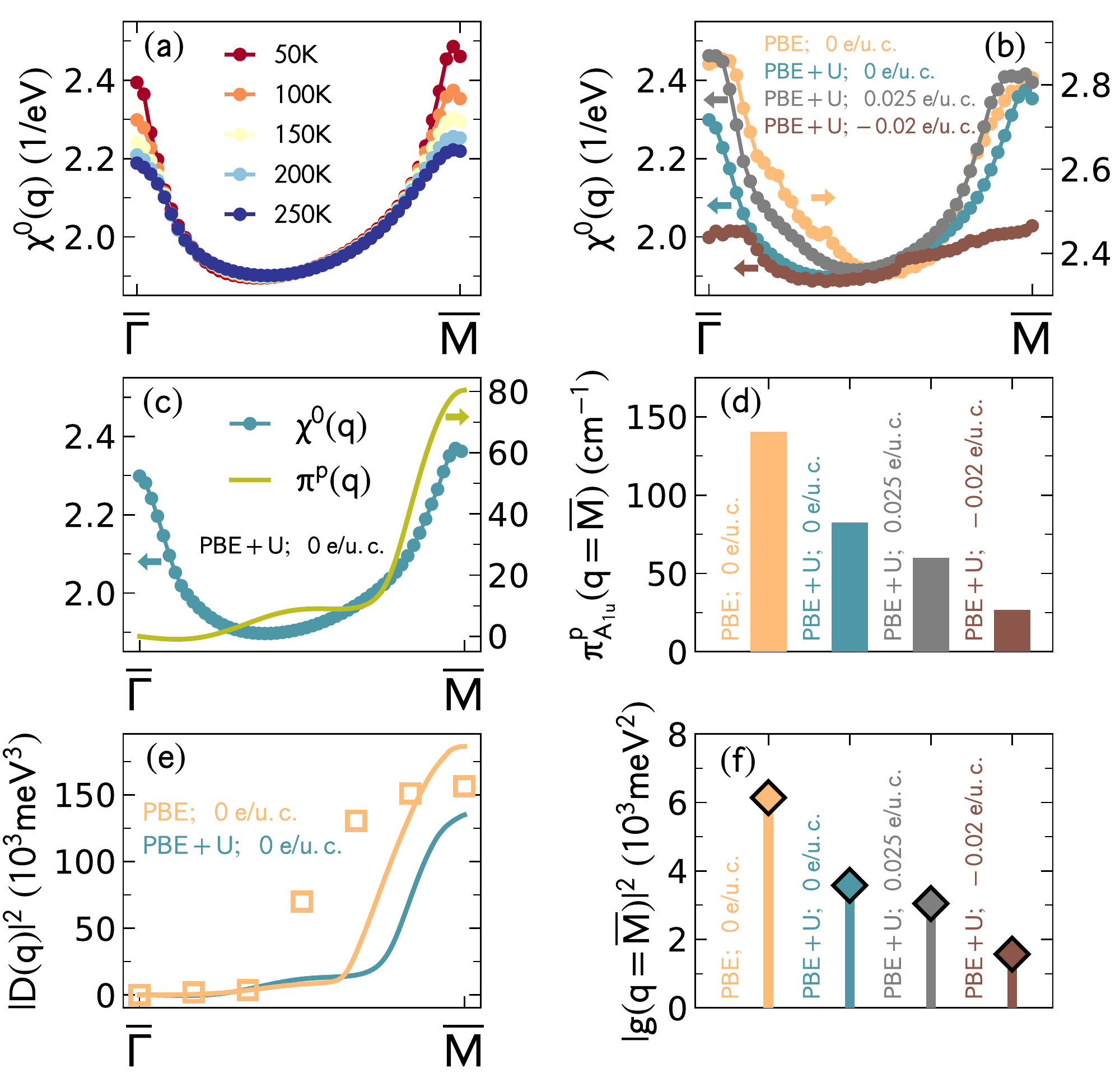}
\caption{(a) Charge correlation function $\chi^0(\mathbf{q})$ as obtained with PBE$+$U along the $\overline{\Gamma}-\overline{\mathrm{M}}$ path and as a function of temperature. (b) Comparison between $\chi^0(\mathbf{q})$ calculated with PBE and PBE$+$U and for different carrier dopings. (c) Partially-screened phonon self-energy $\pi^{p}_{\nu}(\mathbf{q})$ of the CDW mode ($\nu=A_{\rm 1u}$) along $\overline{\Gamma}-\overline{\mathrm{M}}$ (left y-axis) compared with $\chi^0(\mathbf{q})$ (left y-axis). (d) $\pi^{p}_{\nu}(\mathbf{q})$ for $\nu=A_{\rm 1u}$ and $\mathbf{q}=\mathrm{\overline{M}}$ as obtained with PBE and PBE$+$U, as well as with different dopings. (e) Effective deformation potential $|D_{\nu}(\mathbf{q})|^2$ for the CDW mode along $\overline{\Gamma}-\overline{\mathrm{M}}$ path obtained with PBE and PBE$+$U, extracted from cDFPT. For comparison we show the PBE results obtained with the standard DFPT methods\,\cite{qe,baroni01} for extracting the EPC matrix elements (orange squares). (f) Extracted effective $|g_{\nu}(\mathbf{q})|^2$ for $\nu=A_{\rm 1u}$ and $\mathbf{q}=\mathrm{\overline{M}}$ as obtained with PBE and PBE$+$U, as well as with different dopings. Results obtained in panels (b)-(f) are for $T=100$\,K. 
}
\label{fig2}
\end{center}
\end{figure}
%%%%%%%%%%%%%%%%%%%%%%%%%%%%%

Electron-lattice contribution to the dynamical matrix consists of static phonon self-energy $\pi_{\nu}(\mathbf{q})$, while the phonon frequency renormalization of mode $\nu$ can be written as $\omega_{\mathbf{q}\nu}^2=(\omega_{\mathbf{q}\nu}^{b})^2+2\omega_{\mathbf{q}\nu}^b\pi_{\nu}(\mathbf{q})$, where $\omega^b_{\mathbf{q}\nu}$ and $\omega_{\mathbf{q}\nu}$ are bare and renormalized phonon frequencies due to EPC\,\cite{baroni01,giustino17}. If one assumes a weak coupling theory (i.e., no high-order EPC terms as it is the case in DFPT)\,\cite{varma83} and the effective EPC function for which the electronic degrees of freedom are averaged out, i.e.,  $g_{\nu}^{nm}(\mathbf{k},\mathbf{q})\approx g_{\nu}(\mathbf{q})$, the phonon self-energy can be written as $\pi_{\nu}(\mathbf{q})\approx|g_{\nu}(\mathbf{q})|^2\chi^0(\mathbf{q})$, where $\chi^0(\mathbf{q})$ is the bare charge correlation function\,\cite{zhu15,zhou20}. In that way we can juxtapose the roles of Fermi surface nesting via $\chi^0(\mathbf{q})$ and EPC via $|g_{\nu}(\mathbf{q})|^2$.

Figure\,\ref{fig2}(a) shows the calculated $\chi^0(\mathbf{q})$ for PBE$+$U along $\overline{\Gamma}-\overline{\mathrm{M}}$, where only intraband and interband electronic transitions between two Se-$p$ and one Ti-$d$ states in the original $1\times 1$ cell are considered. Temperature dependence arises only from the Fermi-Dirac factors, i.e., from the electrons. There are two pronounced peaks in $\chi^0(\mathbf{q})$, i.e., at $\mathbf{q}=\overline{\Gamma}$ and $\mathbf{q}=\overline{\mathrm{M}}$, that additionally show moderate temperature dependence. $\chi^0(\mathbf{q})$ at $T=100$\,K with and without U are compared in Fig.\,\ref{fig2}(b). Even though low-energy band structure is considerably modified by the inclusion of U, the structure of the peak at $\mathbf{q}=\overline{\mathrm{M}}$ is almost unchanged. Hence, the electron-correlation-induced hardening of the CDW mode at $\mathbf{q}=\overline{\mathrm{M}}$ [Fig.\,\ref{fig1}(c)] is probably driven by modifications of EPC, rather than the nesting properties (as in Peierls scenario).
In addition, we show how small electron and hole dopings affect $\chi^0(\mathbf{q})$. That is, populating the Ti-$d$ valley at M with excess charge carriers does not lead to significant modifications of $\chi^0(\mathbf{q})$, while the depopulation of both electron and hole pockets at $\Gamma$ and M alters considerably the phase space for electronic transitions (see also Fig.\,S1\ in SM\,\cite{SM}).

Further, we provide a comparison between $\chi^0(\mathbf{q})$ and the phonon self-energy $\pi^{p}_{\nu}(\mathbf{q})$ in order to extract the effective EPC function $|g_{\nu}(\mathbf{q})|^2$ and deformation potential $|D_{\nu}(\mathbf{q})|^2=2\omega_{\nu}|g_{\nu}(\mathbf{q})|^2$ for the $\nu=A_{\rm 1u}$ mode. The partially-screened phonon self-energy $\pi^{p}_{\nu}(\mathbf{q})$ accounting only for relevant electron transitions between low-energy Ti and Se bands is obtained by means of constrained DFPT (cDFPT)\,\cite{novko2020a,berges20,nomura15} (see also Sec.\,S1 in SI). The results are reported in Figs.\,\ref{fig2}(c)-(f). $\pi^{p}_{A_{1u}}(\mathbf{q})$ shows considerable momentum dependence along $\overline{\Gamma}-\overline{\mathrm{M}}$, as well as sensitivity to inclusion of U and excess charge carriers. More remarkably, we show that the effective EPC function is momentum dependent and, thus is a prime driving force of the $\mathbf{q}=\overline{\mathrm{M}}$ Kohn anomaly and the accompanying lattice instability. Moreover, it is evident from Fig.\,\ref{fig2}(f) that the inclusion of the proper electron correlations via U reduces the EPC strength and deformation potential (see also Fig.\,S3 in SM\,\cite{SM}). One can therefore conclude that the momentum-dependent EPC plays a more dominant role than the Fermi surface nesting for the CDW formation in TiSe$_2$, similarly to the case of NbSe$_2$\,\cite{johannes06,calandra09,zhu15}. This is in line with early theoretical considerations\,\cite{yoshida80}, however, contrary to more recent models, where $\mathbf{q}$-independent EPC is considered\,\cite{monney15}, as well as to what was recently concluded in time-resolved ultrafast electron diffraction study\,\cite{otto21}. Note that temperature dependence of the Kohn anomaly is, nevertheless, ruled by the entropy of the electron subsystem entering via $\chi^0(\mathbf{q})$. 

The results also demonstrate that modifications of nesting properties and EPC strength of the CDW mode both collaborate in the carrier-induced melting process of the CDW. Theoretical modeling of the temperature-charge phase diagram in TiSe$_2$ is important for comprehending the following experimental observations: the competition of the superconductive and CDW orders in bulk TiSe$_2$ as a function of doping\,\cite{morosan06,qian07}, the corresponding melting of the CDW order in 1L and bulk\,\cite{kolekar18,watson20}, and generally higher values of $T_{\rm CDW}$ in supported 1L TiSe$_2$, where unavoidable Se vacancies might be present leading to the intrinsic electron doping\,\cite{sugawara16,jia18,watson20}.

%%%%%%%%%%%%%%%%%%%%%%%%%%%%%
%%%%%%%%%%%%%%FIGURE%%%%%%%%%
%%%%%%%%%%%%%%%%%%%%%%%%%%%%%
\begin{figure*}[!t]
\begin{center}
\includegraphics[width=0.8\textwidth]{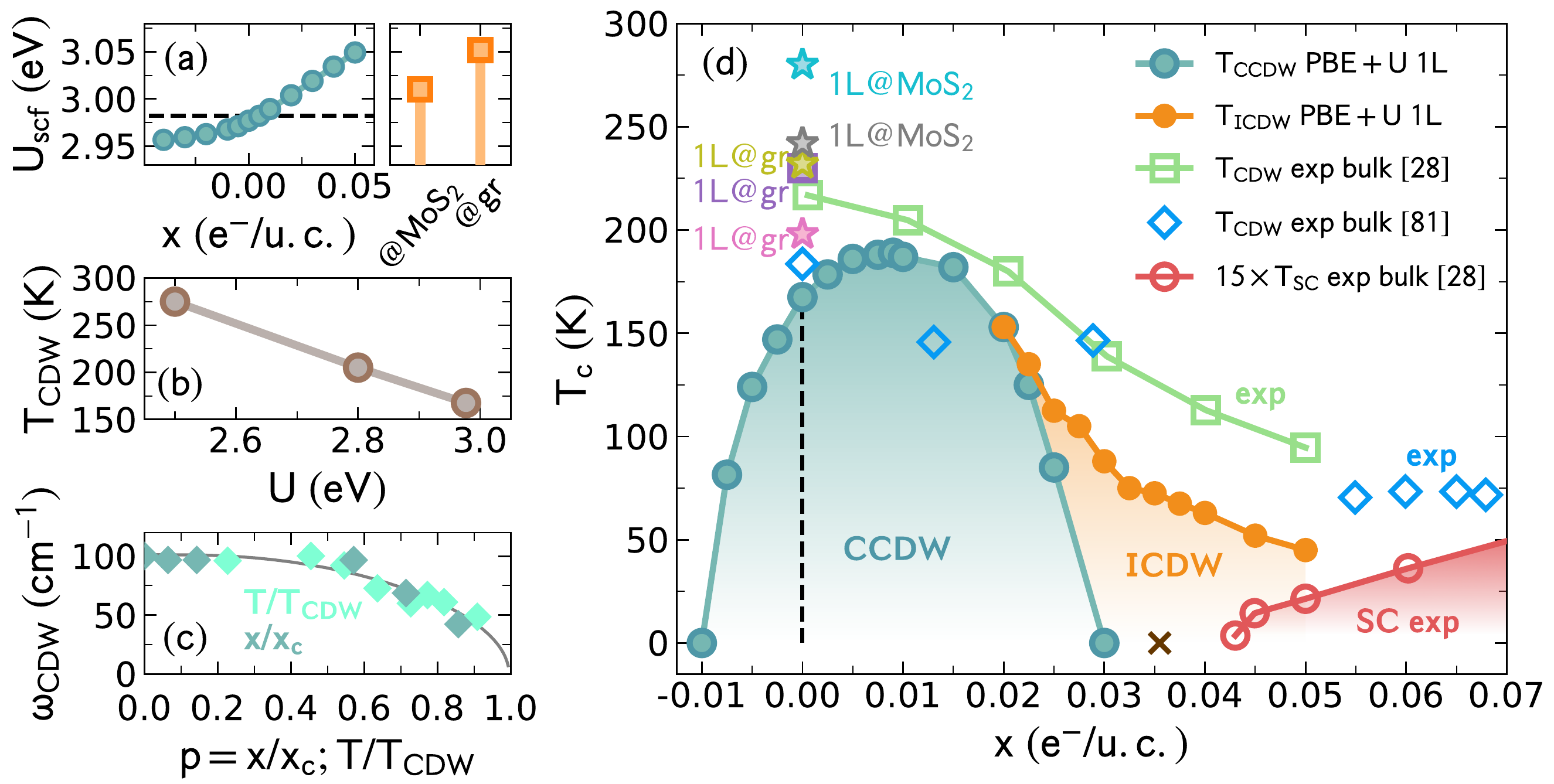}
\caption{(a) Self-consistently obtained Hubbard parameter U as a function of electron concentration and when 1L TiSe$_2$ is adosrbed on 3L MoS$_2$ and 3L graphite. (b) Transition temperature $T_{\rm CDW}$ for arbitrarily chosen Hubbard parameters U. (c) Softening of the CDW amplitude mode obtained with PBE$+$U as a function of electron doping x and temperature. The results are fitted to $a\tanh({b\sqrt{p-1}})$, where $a=101.2$\,cm$^{-1}$, $b=1.42$, $\mathrm{p=x/x_c}$ or $\mathrm{p}=T/T_{\rm CDW}$, $\mathrm{x_c=0.035\,e^-/u.c.}$, and $T_{\rm CDW}=219$\,K. (d) Temperature-charge phase diagram of 1L TiSe$_2$ as obtained with PBE$+$U for commensurate and incommensurate CDW phases (blue and orange circles, respectively). Experimental values of $T_{\rm CDW}$ for 1L TiSe$_2$ on graphite/graphene are depicted with pink star\,\cite{kolekar18}, purple square\,\cite{kolekar17}, and green star\,\cite{chen15}, while for 1L on MoS$_2$ with grey\,\cite{kolekar18} and blue stars\,\cite{kolekar17}. Experimental phase diagrams for bulk are from Refs.\,\cite{morosan06} and \cite{kogar17b}. Brown cross marks the quantum critical point as estimated in Refs.\,\cite{disalvo76} and \cite{jaouen19}.
}
\label{fig3}
\end{center}
\end{figure*}
%%%%%%%%%%%%%%%%%%%%%%%%%%%%%

Figure \ref{fig3}(a) depicts the self-consistently obtained U values as a function of hole and electron doping concentrations. In addition, values of U for Ti atoms are provided when 1L TiSe$_2$ is adsorbed on 3L MoS$_2$ and 3L graphite (see Fig.\,S8 in SM for more details on the structures\,\cite{SM}), as it is often the case in the experiments\,\cite{kolekar18,sugawara16,watson20}. From these results it is evident that modifications of charge concentrations and dielectric environment does not provide sufficient alterations of Hubbard parameter U that could in turn explain discrepancies of $T_{\rm CDW}$ for 1L deposited on graphene/graphite and MoS$_2$\,\cite{kolekar17,kolekar18}. In fact, as demonstrated in Fig.\,\ref{fig3}(b), much higher modifications of U are necessary in order to increase $T_{\rm CDW}$ by about $\sim 10$\,K.
Screening of the EPC function $g_{\nu}(\mathbf{q}=\mathrm{\overline{M}})$ due to the presence of the substrate $\varepsilon_{\rm S}$ might also be ruled out, since the corresponding screened Coulomb interaction is $\mathrm{W}_{\rm S}=[1+(1-\varepsilon_{\rm S})/(1+\varepsilon_{\rm S})e^{-2 |{\bf q}| d}]\mathrm{V}$ (where d is distance between 1L and substrate, while V is the bare Coulomb interaction)\,\cite{despoja19} so that the screening due to substrate is 1, i.e., $\mathrm{W_{S}=V}$, at the edge of the BZ (i.e., for large $\mathbf{q}$).
On the other hand, different concentrations of the Se vacancies might explain the discrepancies between $T_{\rm CDW}$ reported  in various experiments\,\cite{zhou20}, as well as different lengths of the Ti-Se bonds, which provide considerable modifications to the $T_{\rm CDW}$ (see Fig.\,\ref{fig1}(d), corresponding discussion, as well as Ref.\,\cite{fu16}).

As depicted in Fig.\,\ref{fig3}(c), the doping considerably affects the CDW amplitude mode (i.e., for $T<T_{\rm CDW}$) that softens towards $\mathrm{x=x_c\approx0.035\,e^-/u.c.}$ in the same manner as the temperature-induced softening towards $T_{\rm CDW}$. Note that our results for the critical doping $\mathrm{x_c}$ agrees very well with the experimental estimation of the quantum critical point\,\cite{disalvo76,jaouen19}. The analysis on momentum-dependent electron-phonon coupling reported above reveals that the doping-dependent softening of the amplitude mode comes from the reduction of the electron-phonon coupling strength [see Fig.\,\ref{fig2}(f)], as it was speculated in Ref.\,\cite{barath08}.

Temperature-charge phase diagram as obtained with PBE$+$U is reported in Fig.\,\ref{fig3}(d), along with the experimental values for bulk and 1L\,\cite{morosan06,chen15,kolekar18,kolekar17,kogar17b}. Experiments have thus far only provided the values of $T_{\rm CDW}$ for electron doping\,\cite{morosan06,watson20}, while here we calculate the full phase diagram, which includes also the CDW melting via hole carriers.
Interestingly, maximum value of $T_{\rm CDW}$ is obtained for $\mathrm{x=0.009\,e^-/u.c.}$, which is closer to the experimental values of 1L on graphite\,\cite{kolekar18}, indicating that supported 1L samples are doped. In Ref.\,\cite{jia18} the estimated doping for TiSe$_2$ on graphene is $\mathrm{x=0.02\,e^-/u.c.}$, for instance. A dome-like structure analogous to the superconductivity phase diagram\,\cite{morosan06} is obtained for the $\mathbf{q}=\overline{\mathrm{M}}$ instability, corresponding to the commensurate $2\times2$ CDW (CCDW) phase (blue circles). Starting at around $\mathrm{x=0.02\,e^-/u.c.}$, we also observe a phonon instabilities away from the $\overline{\mathrm{M}}$ point (see Fig.\,S9\,\cite{SM}) with a higher critical temperatures (orange points), which forms the incommensurate CDW (ICDW) phase\,\cite{joe14,li16,kogar17b,jaouen19,chen19,li19}. For these doping concentrations, the occupation of the electron Ti-$d$ pocket at the M point increases, which renders the perfect $\mathbf{q}=\overline{\mathrm{M}}$ electron-hole transitions less, while transitions with $\mathbf{q}<\overline{\mathrm{M}}$ more probable.
Interestingly, for hole dopings we do not obtain the ICDW phase. By combining these results we get a very good agreement with the experiments\,\cite{morosan06,kogar17b}, showing that the total experimental phase diagram consists of the CCDW and ICDW regimes. Further, at the point where the CCDW is completely melted ($\mathrm{x_c\approx 0.03-0.04\,e^-/u.c.}$; see brown cross for the experimental estimation of the quantum critical point\,\cite{disalvo76,jaouen19}), the superconducting phase emerges\,\cite{morosan06}, coexisting with the ICDW\,\cite{li16}. This theoretically confirms already speculated relationship between the incommensurate ordered phases and superconductivity, as observed in various transition metal dichalcogenides. Note that standard PBE calculations without Hubbard corrections, on the other hand, grossly overestimates the point of the CDW melt and the overall phase diagram\,\cite{chen18,guster18}. As mentioned, the anharmonic corrections improve the CDW melting picture compared to the harmonic PBE results, however, when it comes to the electron-doping properties described above ($T$-doping phase diagram, doping-induced softening of the CDW amplitude mode, and the ICDW phase) the electronic degrees of freedom along with the electron-phonon coupling seemed to be more relevant compared to the phonon entropy (e.g., in Ref.\,\cite{zhou20} no signs of these properties are observed in electron-doped TiSe$_2$ where $\mathrm{x=0.05\,e^-/u.c.}$ was used).

% Electrical transport measurements under electric-field-induced doping; superconductivity and ICDW under doping - li16

% ICDW under pressure, XRD study - joe14

% STS study, DW and ICDW in TiSe2, Cu doping - yan17

% XRD, ICDW in phase diagram, Cu doping - kogar17b

%incommensurate CDW (ICDW) \cite{joe14,li16,kogar17b,yan17}

Simple qualitative condition for the stability of the CDW ground state can be written as $4|g(\mathbf{q})|^2/\omega_{\bf q}-2\mathrm{U}_{\bf q}+\mathrm{V}_{\bf q}\ge 1/\chi^{0}(\mathbf{q})$, where $\mathrm{U}_{\bf q}$ is the Coulomb interaction and $\mathrm{V}_{\bf q}$ is the screened exchange interaction\,\cite{chan73,rossnagel11}. Here we have demonstrated that considerable momentum-dependent EPC $|g(\mathbf{q})|^2$ and broadened singularity in $\chi^0(\mathbf{q})$ stabilizes the CDW order in TiSe$_2$, in accordance with the above condition. Without proper description of $\mathrm{U}_{\bf q}$ and $\mathrm{V}_{\bf q}$, as it is the case for semi-local exchange-correlation functionals such as PBE, the calculated transition temperature is an order of magnitude larger than the experimental one, i.e., $T^{\rm PBE}_{\rm CDW}\gg T^{\rm exp}_{\rm CDW}$. The present study highlights the role of on-site Coulomb (Hubbard) interaction U that makes the CDW ``less stable'', hardens the Kohn anomaly, and greatly improves the agreement with experimental observations, e.g., $T^{\rm PBE+U}_{\rm CDW}\approx T^{\rm exp}_{\rm CDW}$. Importance of the screened exchange interaction $\mathrm{V}_{\bf q}$ was already demonstrated\,\cite{hellgren17,zhou20}, so one can speculate that its inclusion along with U might lift the agreement with the experiments even more (e.g., increase the PLD and the CDW gap). Recent ARPES experiments, in addition, suggest strong-coupling theory of CDW in TiSe$_2$\,\cite{watson20}, where larger distortions are expected and therefore nonlinear, high-order electron-phonon terms and anharmonicity might be important\,\cite{varma83,yoshiyama86,flicker15,zhou20}.
%, along with the dynamical effects beyond the adiabatic approximation\,\cite{yoshiyama86}.
All these renders the CDW phase in TiSe$_2$ a highly unconventional, where all degrees of freedom are actively involved, while phonons have a leading role.

\begin{acknowledgments}
We gratefully acknowledge useful discussions with R. Bianco, J. Berges, S. Ponc\'e, E. Cappelluti, E. Tuti\v{s}, O. S. Bari\v{s}i\'c, and J. Krsnik. D.N. acknowledges financial support from the Croatian Science Foundation (Grant no. UIP-2019-04-6869) and from the European Regional Development Fund for the ``Center of Excellence for Advanced Materials and Sensing Devices'' (Grant No. KK.01.1.1.01.0001). Part of the computational resources were provided by the DIPC computing center.
\end{acknowledgments}

%\section{Data Availability Statement}
%The data underlying this study are openly available in figshare at \href{dx.doi.org/10.6084/m9.figshare.21353883}{dx.doi.org/10.6084/m9.figshare.21353883}.

%\begin{suppinfo}
%More information on numerical calculations, constrained density functional perturbation theory, electronic structures of $1\times1$ and $2\times2$ cells, 1D anharmonic model, periodic lattice distortions, incommensurate phonon instabilities, and structures of supported 1L TiSe$_2$.\\
%\end{suppinfo}

\bibliography{tise2}

\end{document}